\documentstyle[aps,floats,pra]{revtex}

\input epsf

\tighten

\begin{document}

\draft

\title{Orchestrating an NMR quantum computation: the $N=3$
Deutsch-Jozsa algorithm.}  

\author{ David~Collins,$\mbox{}^{1}$ 
         K.~W.~Kim,$\mbox{}^{1,}$\cite{kiwook} 
         W.~C.~Holton,$\mbox{}^{1}$
         H. Sierzputowska-Gracz,$\mbox{}^{2}$ and 
         E. O. Stejskal$\mbox{}^{3}$ }

\address{ $\mbox{}^{1}$ Department of Electrical and Computer
Engineering, Box 7911, North Carolina State University, Raleigh, North
Carolina 27695-7911 \\ 
          $\mbox{}^{2}$Department of Biochemistry, Box
7622, North Carolina State University, Raleigh, North Carolina
27695-7622 \\ 
          $\mbox{}^{3}$Department of Chemistry, Box 8204, North
Carolina State University, Raleigh, North Carolina 27695-8204}

\date{\today}
      
\maketitle    

\begin{abstract}
A detailed description of the development of a three qubit NMR
realization of the Deutsch-Jozsa algorithm [Collins et.al., Phys.\
Rev.\ A {\bf 62} 022304 (2000)] is provided. The theoretical and
experimental techniques used for the reduction of the algorithm's
evolution steps into a sequence of NMR pulses are discussed at
length. This includes the description of general pulse sequence
compilation techniques, various schemes for indirectly coupled gate
realizations, experimental pulse parameterization techniques and
bookkeeping methods for pulse phases.
\end{abstract}


\section{Introduction}

Nuclear magnetic resonance (NMR) spectroscopy has become well
established as a tool for exploratory experimental investigations of
quantum computation. Fundamental gates, algorithms, error correction
and other issues in quantum information theory have been demonstrated
at the level of a few qubits using room temperature, solution state
NMR
systems.\cite{chuang,jones,linden,chuang2,cory1,cory2,knill,cory3,madi,marx,dorai1,collins2,kimlee,dorai2,lieven1,zhu,lieven2,somaroo,knill2,nielsen,nelson,fang,chuang3,yannoni,jones2,lindenkupce}

Recently we successfully implemented the $N=3$ case of the
Deutsch-Jozsa algorithm using room temperature solution state
NMR.\cite{collins2} This implementation was characterized by the
choice of qubits, namely the $^{13}\mbox{C}$ nuclei of alanine, and
the resulting methods for processing the quantum information. The
homonuclear nature of the qubits (henceforth qubit and spin are used
interchangeably) demands qubit selective pulses throughout the
realization, thus greatly complicating the experimental effort in
comparison to a heteronuclear system. Furthermore the arrangement of
couplings between qubits in alanine is such that the only reliable
method for processing quantum information is along a chain of qubits,
whose pattern of couplings is effectively $A - B - C$ (i.e.\ the $A -
C$ coupling is too weak). This demands ``indirect'' gate realizations
using swapping techniques via intermediate spins~\cite{lloyd} and is
more difficult than ``direct'' realizations, possible in the fully
coupled case. Although these restrictions are artificial at the level
of a few qubits there is reason to believe that they will become
important as the number of qubits
increases~\cite{lindenkupce}. Therefore the ability to implement the
associated information processing techniques will become crucial;
hence our choice of this restrictive qubit system.

Even at the low level of three qubits the resulting experiments were
sufficiently complicated to motivate the development and use of a host
of simplification techniques and experimental methods to construct the
appropriate pulse sequence. Much of the literature bypasses or only
briefly describes such details, which will become increasingly important
as the number of qubits in NMR quantum computation implementations
grows.  The purpose of this article is to describe in detail the
construction of an NMR realization from the theoretical algorithm
through to the final spectrometer output and the algorithm
result. Section~\ref{sec:dj} briefly describes the Deutsch-Jozsa
problem and algorithm and the initial decomposition of the algorithm's
evolution stage into fundamental quantum
gates. Section~\ref{sec:qctheory} covers the theoretical development
of initialization schemes, pulse sequences and readout
schemes. General pulse sequence compilation techniques and methods for
constructing indirectly coupled gate realizations are presented
here. Section~\ref{sec:exp} describes the experimental details and
includes pulse parameterization, bookkeeping methods for pulse phases,
construction of appropriate evolution delays, results and possible
sources of error.

\section{Deutsch-Jozsa Algorithm}
\label{sec:dj}

\begin{figure}[ht] 
  \epsfxsize=4in
  \centerline{\epsffile{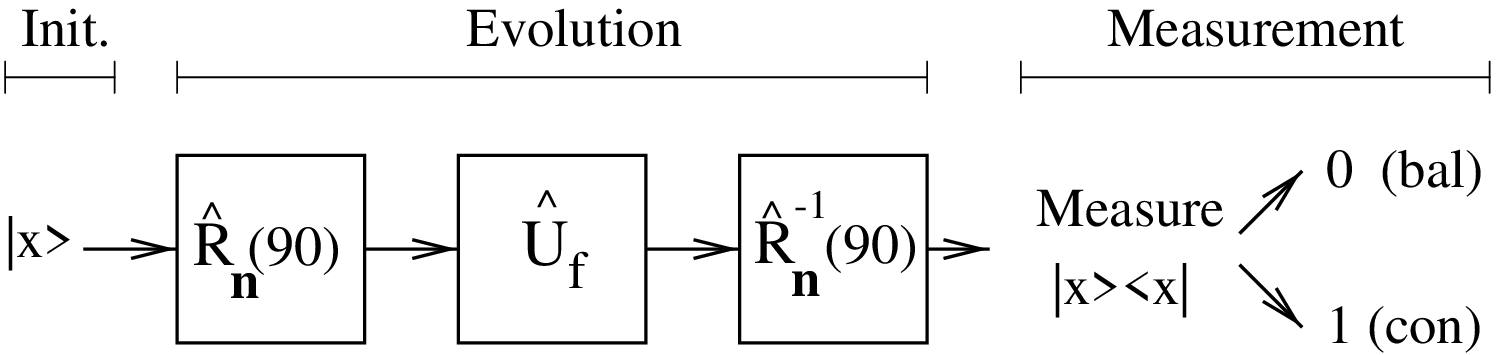}}
  \vspace{2ex}
  \caption{Deutsch-Jozsa algorithm. Initialization to the state: 
$\left| x \right> \equiv \left| x_{N-1} \right> ... \left| x_0 \right>$
where $x_i \in \{ 0,1 \}$. $\hat{R}_{{\bf n}} (90)$ rotates each qubit
through $90^\circ$ about 
${\bf n} = \cos{\phi}\hat{\bf x} +\sin{\phi}\hat{\bf y}.$
 For a single qubit 
$\hat{R}_{{\bf n}}(\theta) := e^{-i{\bf n}.\hat{\bf \sigma} \theta/2}$ 
where 
$\hat{\bf n}.\hat{\bf \sigma} = n_x \hat{\sigma}_x + 
                                n_y \hat{\sigma}_y + 
                                n_z \hat{\sigma}_z$.
The $f$-controlled gate, $\hat{U}_f \left| x \right>
:= \left( -1\right)^{f(x)} \left| x \right>$, evaluates $f$. The
expectation value of $\left| x \right>\left< x \right|$ on the output
state answers the problem conclusively. }  
\label{pic:djscheme}
\end{figure}

The Deutsch problem~\cite{deutsch,cleve} addresses a global property
of certain binary valued functions $f:\{0,1\}^N \rightarrow
\{0,1\}$. In particular, a function is {\em balanced} if it returns
$0$ as many times as $1$ after evaluation over its entire range. Given
any function which is either balanced or constant the problem is to
determine its type. Classical algorithms which rely on repeatedly
evaluating the function for various arguments require a number of
evaluations which grows exponentially with $N$ to answer the problem
with certainty~\cite{cleve}. However, a quantum algorithm requires a
single evaluation of $f$ and solves the problem with
certainty~\cite{deutsch,cleve,collins}. In its simplest rendition the
algorithm needs an $N$ qubit register~\cite{collins} and follows the
scheme illustrated in Fig.~\ref{pic:djscheme}.

The ease of implementation relative to other quantum algorithms has
made the Deutsch-Jozsa algorithm a natural candidate for experimental
NMR demonstrations of features of quantum computation. The $N~=~1$
case, using the Cleve version~\cite{cleve}, was the first quantum
algorithm implemented experimentally~\cite{chuang,jones}. This was
followed by the $N=2$ case~\cite{linden,dorai1}. However, the
algorithm is uninteresting from a quantum computational perspective
for $N~<~3$, in which case the essential operations could be
implemented classically~\cite{collins}. Recently the modified version
(see Fig.~\ref{pic:djscheme}) of the $N~=~3$ algorithm has been
implemented for various balanced
functions~\cite{collins2,kimlee,dorai2}. Finally, limited balanced
function cases at the $N~=~4$ level have been tested using the Cleve
version~\cite{marx}.

\subsection{Gate Decomposition}

The heart of the Deutsch-Jozsa algorithm is the function evaluation
step, 
\begin{equation}
  \hat{U}_f \left| x \right> := 
                       \left( -1\right)^{f(x)} \left| x \right>,
 \label{eq:ufdef}
\end{equation}
and the bulk of the effort in any experiment will be devoted to its
realization. This is initiated by decomposing $\hat{U}_f$ into a
product of fundamental one and two qubit operations which are amenable
to physical realization. In general single bit rotations (one qubit
operations) and controlled-NOT gates (two qubit operations) are
satisfactory.\cite{barenco} In this scheme, the hope is to be able to
devise physical realizations for the fundamental gates and concatenate
these to recreate $\hat{U}_f$. The definition of $\hat{U}_f$,
Eq.~(\ref{eq:ufdef}), does not immediately provide such a
decomposition. One approach for arriving at a fundamental gate
decomposition begins by representing any admissible function via power
series expansions in the argument bits, $x_i$ where $i=0,1,2$. Such
expansions terminate since $x_i^2 = x_i$ for $x_i \in \{ 0,1 \}$,
leaving at most a cubic term for $N=3$. However, it can be shown that
for balanced and constant functions the cubic term disappears. Thus
for $N=3$ any admissible function can be expanded as:
\begin{equation}
  f(x_2,x_1,x_0) = \bigoplus_{i>j \geq 0}^2 a_{ij} x_i x_j 
                   \bigoplus_{i=0}^2 a_i x_i \oplus a
  \label{eq:seriesdecomp}
\end{equation}
where addition is modulo 2 and $a_{ij},a_i,a \in \{ 0,1 \}$.  Therefore
\begin{equation}
 \hat{U}_f \left| x_2x_1x_0 \right> = \prod_{i>j \geq 0}^2 
                              \left( -1 \right)^{a_{ij} x_i x_j}
                              \prod_{k=0}^2 \left( -1 \right)^{a_k x_k}
                              \left( -1 \right)^a
                            \left| x_2x_1x_0 \right>.
\end{equation}
The constant term merely provides an irrelevant overall phase factor
and can be ignored. The quadratic and linear terms motivate the
definitions of the {\em quadratic term gates} and {\em linear term
gates}:
\begin{eqnarray}
 \hat{U}^{ij}_{\mbox{\tiny QUAD}}\left| x \right> & := & 
                         \left( -1\right)^{x_ix_j} \left| x \right>
                         \mbox{ and} \label{eq:quadgate}\\
 \hat{U}^i_{\mbox{\tiny LIN}}\left| x \right> & :=  & 
                         \left( -1\right)^{x_i} \left| x \right>,
                         \label{eq:lingate}
\end{eqnarray}
leaving
\begin{equation}
        \hat{U}_f = \prod_{i>j \geq 0}^2 \left( 
                      \hat{U}^{ij}_{\mbox{\tiny QUAD}} \right)^{a_{ij}}
                    \prod_{k=0}^2 \left( \hat{U}^{k}_{\mbox{\tiny LIN}} 
                                                        \right)^{a_{k}}
 \label{eq:fdecomp}
\end{equation}
where the order of operation reads from right to left. However,
quadratic and linear term gates all commute and any rearrangement is
permitted in this decomposition. Decompositions of these in terms of
fundamental gates are
\begin{eqnarray}
 \hat{U}^i_{\mbox{\tiny LIN}} & = & \hat{R}^i_{\hat{\bf z}}(180^\circ) 
                                    \mbox{ \hspace{3em} and}
 \label{eq:lingateseq} \\
 \hat{U}^{ij}_{\mbox{\tiny QUAD}} & = &
                                \hat{R}^j_{-\hat{\bf y}}(90^\circ)\,
                                \hat{U}^{ij}_{\mbox{\tiny CNOT}}\,
                                \hat{R}^j_{\hat{\bf y}}(90^\circ)
 \label{eq:quadgateseq}
\end{eqnarray}
where $\hat{U}^{ij}_{\mbox{\tiny CNOT}}$ is a controlled-NOT gate with
control $i$ and target $j$ and $\hat{R}^i_{\hat{\bf n}}(\theta)$ is
the operator representation of a rotation of spin $i$ about axis
$\hat{\bf n}$ through angle $\theta$.

\subsection{Function Classification}

There are $2^N \choose{2^{N-1}}$ balanced functions with $N$ bit
arguments and for $N=3$ this gives $70$ distinct
possibilities. However, certain distinct balanced functions are such
that the operations required to realize $\hat{U}_f$ are related by
permutations of the argument qubits. For example, permuting the qubit
indices using the rule $1 \leftrightarrow 2$ transforms the operation
required to realize $f(x_2,x_1,x_0) = x_2$ to that for $g(x_2,x_1,x_0)
= x_1$. The physical difference between the realization for these
cases is minor in comparison to that for a case which is unrelated in
this fashion; an example is $h(x_2,x_1,x_0) = x_2 \oplus x_1$. Thus it
makes sense to classify the balanced functions according to the
similarity of their power series expansions under permutations of
argument indices; this is mirrored by an equivalent classification of
the operations required for function evaluation under qubit
permutations. Within each class the arrangement of linear and
quadratic term gates differ only by permutations of their qubit
indices and this reduces the number of essentially different
experimental realizations needed to demonstrate the algorithm. It
emerges that there are ten classes of balanced functions; a
representative of each is provided in Table~\ref{tab:foutput}. All
possible quadratic term gates are required for the classes represented
by $f_9$ and $f_{10}$ and all possible linear term gates are required
for $f_3$. Thus realization of the algorithm for one representative
from each class requires the construction of all possible linear and
quadratic term gates, which could be used for realization of any other
case. The experiment will be limited to one function from each
balanced class plus the constant functions.

\begin{table}[ht]
\begin{tabular}{ll}
  Representative, $f$   & $\hat{\rho}_f$  \\ \hline
  Constant: &            \\
  $f_{const} = 0$   & $-\hat{I}^2_x - \hat{I}^1_x - \hat{I}^0_x$ \\ 
  Balanced: &            \\
  $f_1 = x_2$                  & $+\hat{I}^2_x - 
                                   \hat{I}^1_x - 
                                   \hat{I}^0_x$ \\
  $f_2 = x_2 \oplus x_1$       & $+\hat{I}^2_x + 
                                   \hat{I}^1_x - 
                                   \hat{I}^0_x$ \\
  $f_3 = x_2 \oplus x_1 
             \oplus x_0 $      & $+\hat{I}^2_x + 
                                   \hat{I}^1_x + 
                                   \hat{I}^0_x$ \\
  $f_4 = x_2 x_1 \oplus x_0$   & $-2 \hat{I}^2_x \hat{I}^1_z
                                  -2 \hat{I}^2_z \hat{I}^1_x 
                                   + \hat{I}^0_x$ \\
  $f_5 = x_2 x_1 \oplus x_2 
                 \oplus x_0 $  & $+2 \hat{I}^2_x \hat{I}^1_z
                                  -2 \hat{I}^2_z \hat{I}^1_x 
                                   + \hat{I}^0_x$\\
  $f_6 = x_2 x_1 \oplus x_2 
                 \oplus x_1 
                 \oplus x_0 $  & $+2 \hat{I}^2_x \hat{I}^1_z
                                  +2 \hat{I}^2_z \hat{I}^1_x 
                                  + \hat{I}^0_x$\\
  $f_7 = x_2 x_1 
         \oplus x_1 x_0 
         \oplus x_2 
         \oplus x_1$           & $+2 \hat{I}^2_x \hat{I}^1_z
                                  +4 \hat{I}^2_z \hat{I}^1_x \hat{I}^0_z 
                                  -2 \hat{I}^1_z \hat{I}^0_x$\\
  $f_8 = x_2 x_1 
         \oplus x_1 x_0 
         \oplus x_2 $          & $+2 \hat{I}^2_x \hat{I}^1_z
                                  -4 \hat{I}^2_z \hat{I}^1_x \hat{I}^0_z 
                                  -2 \hat{I}^1_z \hat{I}^0_x$ \\
  $f_9 = x_2 x_1 
         \oplus x_1 x_0 
         \oplus x_2 x_0 $      & $-4 \hat{I}^2_x \hat{I}^1_z \hat{I}^0_z 
                                  -4 \hat{I}^2_z \hat{I}^1_x \hat{I}^0_z 
                                -4 \hat{I}^2_z \hat{I}^1_z \hat{I}^0_x$ \\
  \hspace{-0.4em}$\begin{array}{cl}
     f_{10} =  & x_2 x_1 \oplus x_1 x_0 \oplus x_2 x_0  \\
               & \oplus x_1 \oplus x_0 
   \end{array}$
         &$-4 \hat{I}^2_x \hat{I}^1_z \hat{I}^0_z 
           +4 \hat{I}^2_z \hat{I}^1_x \hat{I}^0_z 
           +4 \hat{I}^2_z \hat{I}^1_z \hat{I}^0_x$ 
 \end{tabular}
  \vspace{2ex}
  \caption{Representatives of each class of admissible functions ($N=3$) 
and the corresponding density operator after the function evaluation 
step.} 
\label{tab:foutput}
\end{table}

\section{Pulse Sequence Development: Theory}
\label{sec:qctheory}

NMR spectroscopy of spin $\frac{1}{2}$ nuclei of appropriate molecules
in solution offers a readily accessible experimental approach to
quantum
computing.\cite{chuang,jones,linden,chuang2,cory1,cory2,knill,cory3,madi}
Any molecule containing three distinguishable, coupled spin
$\frac{1}{2}$ nuclei in an external magnetic field provides the three
qubits needed to solve the $N=3$ Deutsch problem. To a good
approximation the Hamiltonian for a room temperature, solution-state
sample is
\begin{equation}
 \hat{H} =
\sum_{i=0}^2 \frac{\omega_i}{2} \hat{\sigma}^i_z + \frac{\pi}{2}
\sum_{i>j \geq 0}^2 J_{ij} \hat{\sigma}^i_z \hat{\sigma}^j_z,
\label{eq:ham}
\end{equation}
where $\omega_i$ are the Zeeman frequencies, $J_{ij}$ the scalar
coupling constants, and $\hat{\sigma}^i_z$ Pauli
operators~\cite{slichter}. Superscripts label the spins and identify
them with the corresponding argument bits.

\subsection{Gate construction and compilation}

Equations~(\ref{eq:fdecomp})~-~(\ref{eq:quadgateseq}) provide a method
for constructing a real pulse sequence for $\hat{U}_f$ in terms of an
appropriate sequence of single bit rotations and controlled-NOT
gates. Pulse sequences for the latter then enable a practical
realization of the algorithm.  The resulting sequences can be
displayed, following conventional quantum mechanics formalism, as a
sequence of operators with order of application read from right to
left, $\hat{U}_m \cdots \hat{U}_2\,\hat{U}_1$. An alternative, more
familiar to NMR users, is a sequence of pulses, $\left[ U_1 \right] -
\left[ U_2 \right] - \cdots - \left[ U_m \right]$, with order of
application read from left to right. In the room temperature,
solution-state NMR paradigm two types of operations are necessary and
sufficient for gate construction.  These are spin selective rotations
caused by appropriate externally applied RF pulses, and delays during
which the system evolves under only one of the Hamiltonian coupling
terms. The process of pulse sequence construction is initially
simplified by considering spin selective rotations which are idealized
in the sense that they are instantaneous and yet perfectly selective.
An ideal rotation of spin $i$ about axis $\hat{{\bf n}}$ through angle
$\theta$ will be represented by the pulse sequence term $\left[ \theta
\right]^i_n$, corresponding to the operator $\hat{R}^i_{\hat{\bf
n}}(\theta)$. A delay or period of free evolution of duration $t$
under the scalar coupling term $\frac{\pi}{2} J_{ij} \hat{\sigma}^i_z
\hat{\sigma}^j_z$ (no sum over the indices) alone will be represented
by
\begin{equation}
 \hat{U}_{\mbox{\tiny SCAL}}^{ij}(t) := e^{-i\frac{\pi}{2} J_{ij}
\hat{\sigma}^i_z \hat{\sigma}^j_z t }
\end{equation}
or the pulse sequence term $\left[ t \right]^{ij}$. This operation is
unnatural in the sense that some of the scalar coupling terms are
absent. In practice it is more convenient to build this from periods
of free evolution of duration $t$ under all scalar couplings, one of
which will be denoted
\begin{equation}
 \hat{U}_{\mbox{\tiny SCAL}}^{\mbox{\scriptsize tot}}(t) 
        := e^{-i \sum_{j>k \geq 0}^2 \frac{\pi}{2} J_{jk}
\hat{\sigma}^j_z \hat{\sigma}^k_z t }
\end{equation}
or $\left[ t \right]^{\mbox{\scriptsize tot}}$.

Single qubit gates, equivalent to qubit selective rotations, can be
implemented by applying spin selective RF pulses of the appropriate
shape, power and duration~\cite{slichter}. Controlled-NOT gates can be
realized using the rotation and delay method~\cite{chuang,jones,cory2}.
There are two idealized decompositions (i.e.\ in terms of idealized
rotations):
\begin{equation}
 U^{ij}_{\mbox{\tiny CNOT}} = \hat{R}^j_{\bf -\hat{z}}(90^\circ)\;
                              \hat{R}^i_{\bf \pm \hat{z}}(90^\circ)\;
                              \hat{R}^i_{\bf \pm \hat{x}}(90^\circ)\;
                              \hat{U}_{\mbox{\tiny SCAL}}^{ij}
                                           (1/2J_{ij})\;
                              \hat{R}^i_{\bf \pm \hat{y}}(90^\circ)
 \label{eq:cn1}
\end{equation}
or in pulse sequence terms:
\begin{equation}
 \left[ U^{ij}_{\mbox{\tiny CNOT}} \right]
                                 = \left[ 90^\circ \right]^j_{\pm y}
                                 - \left[ 1/2J_{ij} \right]^{ij}
                                 - \left[ 90^\circ \right]^j_{\pm x}
                                 - \left[ 90^\circ \right]^i_{\pm z}
                                 - \left[ 90^\circ \right]^j_{-z}
 \label{eq:cn2}
\end{equation}
where the axes of rotation are given by either all the upper signs or
else all the lower signs.

In order to implement any algorithm, it suffices to simply concatenate
constituent pulse sequences for the required fundamental
gates. However, the difficulty of experimental implementation may be
reduced by various simplification techniques, which can be implemented
efficiently at the algorithm design level via rudimentary, general
schemes, applicable in any physical realization.  These use
\begin{equation}
 \hat{R}^i_{-{\bf \hat{n}}}(\theta)\;
 \hat{R}^i_{\bf \hat{m}}(\phi)\;
 \hat{R}^i_{\bf \hat{n}}(\theta) = \hat{R}^i_{\bf \hat{r}}(\phi)
 \label{eq:rotint}
\end{equation}
where
\begin{equation}
 {\bf \hat{r}} = {\bf \hat{m}} \cos{\theta} 
       - {\bf \hat{n}} \times {\bf \hat{m}} \sin{\theta} 
       + (1- \cos{\theta})({\bf \hat{n}}\cdot{\bf \hat{m}}){\bf \hat{n}}
\end{equation}
and which is easily verified by expansion and multiplication. This
provides two avenues for simplification:
\begin{itemize}
\item[1.] The pulse sequence can be rearranged so that all rotations
about ${\bf \hat{z}}$ follow all other operations. This is possible in
the context of the linear and quadratic term gates because such
rotations commute with periods of evolution under $\hat{H}$ and
because they can be interchanged with rotations about axes in the
transverse plane by applying Eq.~(\ref{eq:rotint}) (here illustrated
for qubit $i$, but applicable for any $i$):
\begin{equation}
 \hat{R}^i_{\cos{\alpha}{\bf \hat{x}} 
          + \sin{\alpha}{\bf \hat{y}}}(\phi)\;
 \hat{R}^i_{\bf \hat{z}}(\theta)
  =  \hat{R}^i_{\bf \hat{z}}(\theta)\; 
       \hat{R}^i_{ \cos{(\alpha - \theta)}{\bf \hat{x}}  
                 + \sin{(\alpha -\theta)}{\bf \hat{y}} }(\phi).
 \label{eq:zrot_rearrange}
\end{equation}
The result is that, all rotations about ${\bf \hat{z}}$ on qubit $i$
can be replaced, after modification of the succeeding rotations on
qubit $i$, by a single rotation about ${\bf \hat{z}}$ on qubit $i$
immediately prior to the measurement step. In the context of solution
state NMR spectroscopy this can be implemented via a qubit dependent
phase shift applied to the acquired data. The need to actively
implement ${\bf \hat{z}}$ rotations is thereby eliminated.

\item[2.] Successive $90^\circ$ rotations (on one qubit and with no
intervening period of evolution under couplings involving this qubit)
about orthogonal axes in the $xy$ plane, ${\bf \hat{m}}$ and ${\bf
\hat{n}}$, can be replaced by a single rotation about an axis in the
$xy$ plane followed by a rotation about ${\bf \hat{z}}$. Again
Eq.~(\ref{eq:rotint}) prescribes:
\begin{equation}
 \hat{R}^i_{\bf \hat{n}}(90^\circ)\;
 \hat{R}^i_{\bf \hat{m}}(90^\circ)
  =  \hat{R}^i_{{\bf \hat{n}}\times {\bf \hat{m}}}(90^\circ)\;
     \hat{R}^i_{\bf \hat{n}}(90^\circ)
\end{equation}
and ${\bf \hat{n}}\times {\bf \hat{m}}$ is oriented along ${\bf
\hat{z}}$. The ${\bf \hat{z}}$ rotation can be interchanged with
all succeeding operations according to the previous scheme. This
reduces the number of rotations requiring active implementation.
\end{itemize}
The number of steps required to carry out these compilation procedures
scales polynomially with the number of gates and thus does not
adversely affect the computational complexity of the
algorithm. Applying these rules to the concatenated sequences for the
Deutsch-Jozsa algorithm yields:
\begin{itemize}
 \item[1.] The linear term gates and other single qubit rotations
about ${\bf \hat{z}}$ can be interchanged with the remaining gates
(with appropriate modifications to their axes of rotation) and
absorbed in the final qubit-dependent phase shifts. These can be
implemented passively via mathematical transformations on the acquired
data and need not be included in the actual physical information
processing stage. The fact that this gives a single manipulation per
qubit means that this strategy does not change the computational
complexity of the problem.
\item[2.] The remaining gates in the sequence are either single qubit
rotations through $90^\circ$ about the ${\bf \hat{x}}$ or ${\bf
\hat{y}}$ axes and scalar coupling gates. For any qubit, the single
qubit rotations which appear between successive scalar coupling gates
involving this qubit can be reduced to at most one rotation about an
axis in the $xy$ plane and at most one rotation about ${\bf \hat{z}}$,
which can be interchanged and absorbed as a phase shift.
\end{itemize}
The result of compilation is a sequence of scalar coupling gates
interspersed with single qubit rotations about axes in the $xy$ plane
and terminated with one qubit dependent rotation about ${\bf \hat{z}}$
per qubit. Thus the sequence of operations involving qubit $i$
(ignoring operations which only involve the remaining qubits) has the
form,
\begin{eqnarray}
 \hat{U}^i(\phi^i,{\mbox{\boldmath $\alpha$}}^i,
                  {\mbox{\boldmath $\theta$}}^i,
                  { \bf t}^i)
 & = & 
 \hat{R}^i_{\bf \hat{z}}(\phi^i) \; 
 \hat{R}^i_{\cos{\alpha^i_{m+1}}{\bf \hat{x}}  
            + \sin{\alpha^i_{m+1}}{\bf \hat{y}}}(\theta_{m+1}) \;
 \hat{U}_{\mbox{\tiny SCAL}}^{ij_{m}}(t^i_{m})\;
 \cdots \nonumber \\
 & & \cdots \;
 \hat{R}^i_{\cos{\alpha^i_2}{\bf \hat{x}}   
          + \sin{\alpha^i_2}{\bf \hat{y}}}
         (\theta_2) \;
 \hat{U}_{\mbox{\tiny SCAL}}^{ij_{1}}(t^i_{1}) \;
 \hat{R}^i_{\cos{\alpha^i_1}{\bf \hat{x}}  
          + \sin{\alpha^i_1}{\bf \hat{y}}}
         (\theta_1)  
 \label{eq:ipulses}
\end{eqnarray}
where $m$ is the number of scalar coupling gates in the sequence, and
${\mbox{\boldmath $\alpha$}}^i := \{ \alpha^i_{m+1}, \ldots,
\alpha^i_1 \}$, ${\mbox{\boldmath $\theta$}}^i := \{ \theta^i_{m+1},
\ldots, \theta^i_1 \}$ and ${\bf t}^i := \{ t^i_m, \ldots, t^i_1 \}$
are arrays of parameters which are determined by the algorithm and
pulse sequence compilation.

The distillation, via the compilation scheme, of all qubit selective
rotations about ${\bf \hat{z}}$ to one rotation per qubit immediately
prior to measurement and the possibility of implementation via
post-acquisition mathematical operations motivates the division of the
pulse sequence into an {\em active section},
\begin{equation}
 \hat{U}^i_{\mbox{\tiny active}}({\mbox{\boldmath $\alpha$}}^i,
                  {\mbox{\boldmath $\theta$}}^i,{\bf t}^i) :=
 \hat{R}^i_{\cos{\alpha^i_{m+1}}{\bf \hat{x}} + 
            \sin{\alpha^i_{m+1}}{\bf \hat{y}}}(\theta_{m+1}) \; 
 \cdots \; 
 \hat{U}_{\mbox{\tiny SCAL}}^{ij_{1}}(t^i_{1}) \; 
 \hat{R}^i_{\cos{\alpha^i_1}{\bf \hat{x}} +
            \sin{\alpha^i_1}{\bf \hat{y}}} (\theta_1),
 \label{eq:active}
\end{equation}
and a {\em passive section},
\begin{equation}
 \hat{U}^i_{\mbox{\tiny passive}}(\phi^i) 
                  :=  \hat{R}^i_{\bf \hat{z}}(\phi^i),
 \label{eq:passive}
\end{equation}
giving
\begin{equation}
   \hat{U}^i(\phi^i,{\mbox{\boldmath $\alpha$}}^i,
                  {\mbox{\boldmath $\theta$}}^i,
                  { \bf t}^i)
 = \hat{U}^i_{\mbox{\tiny passive}}(\phi^i) \;  
   \hat{U}^i_{\mbox{\tiny active}}({\mbox{\boldmath $\alpha$}}^i,
                                   {\mbox{\boldmath $\theta$}}^i,
                                   {\bf t}^i).
\end{equation}
In an NMR realization the active section has to be implemented by
externally applied fields and delays for internal Hamiltonian
evolution. This requires an experiment in which the appropriate pulse
parameters (corresponding to ${\mbox{\boldmath $\alpha$}}^i$ and
${\mbox{\boldmath $\theta$}}^i$) and delays (corresponding to ${\bf
t}^i$) are carefully adjusted so as to achieve the correct evolution.
In contrast the passive section is best implemented mathematically on
the acquired data using the spectrometer's phase shifting capability.

An interesting and useful feature of the active section of the pulse
sequence is that the absolute orientations of its constituent
rotations' axes are less important than their relative orientations
provided that the initial density operator satisfies certain general
conditions. This results from Eq.~(\ref{eq:rotint}) which gives
\begin{equation} 
 \hat{R}^i_{\cos{\alpha}{\bf \hat{x}} 
          + \sin{\alpha}{\bf \hat{y}}}(\theta) 
      = 
 \hat{R}^i_{-{\bf \hat{z}}}(\varphi^i) \;
 \hat{R}^i_{\cos{(\alpha - \varphi^i)}{\bf \hat{x}} 
          + \sin{(\alpha - \varphi^i)}{\bf \hat{y}}}(\theta) \;
 \hat{R}^i_{\bf \hat{z}}(\varphi^i)
\end{equation}
and consequently 
\begin{equation}
 \hat{U}^i(\phi^i,{\mbox{\boldmath $\alpha$}}^i,
                  {\mbox{\boldmath $\theta$}}^i,
                  { \bf t}^i)
  = \hat{U}^i_{\mbox{\tiny passive}}(\phi^i - \varphi^i) \;  
    \hat{U}^i_{\mbox{\tiny active}}({\mbox{\boldmath $\alpha$}}^i
             -{\mbox{\boldmath $\varphi$}^i},
              {\mbox{\boldmath $\theta$}}^i,{\bf t}^i) \;
             \hat{R}^i_{\bf \hat{z}}(\varphi^i).
 \label{eq:ipulses2}
\end{equation}
where ${\mbox{\boldmath $\alpha$}}^i-{\mbox{\boldmath $\varphi$}^i} :=
\{ \alpha^i_{m+1} - \varphi^i, \ldots, \alpha^i_0 - \varphi^i\}$.  The
initial rotation, $\hat{R}^i_{\bf \hat{z}}(\varphi^i)$, can be omitted
whenever the initial state of the system is a mixture of eigenstates
of $\hat{I}_z$ operators. Examples include the thermal equilibrium
state for weakly coupled nuclei and pure (or the equivalent
pseudopure) states such as $\left| x_{N-1} \ldots
x_0\right>$. Henceforth we shall consider this case, so that 
\begin{equation}
 \hat{U}^i(\phi^i,{\mbox{\boldmath $\alpha$}}^i,
                  {\mbox{\boldmath $\theta$}}^i,
                  { \bf t}^i)
  = \hat{U}^i_{\mbox{\tiny passive}}(\phi^i - \varphi^i) \;  
    \hat{U}^i_{\mbox{\tiny active}}({\mbox{\boldmath $\alpha$}}^i
                                   -{\mbox{\boldmath $\varphi$}^i},
                          {\mbox{\boldmath $\theta$}}^i,{\bf t}^i)
 \label{eq:ipulses3}
\end{equation}
and which demonstrates that, for qubit $i$, only the relative
orientations of the axes of the rotations on qubit $i$ are
important. Thus the axis of the initial rotation for each qubit can be
chosen arbitrarily with no need to correlate these between the
different qubits. Thereafter the axes of the succeeding rotations are
fixed. The freedom of choice of the axis of the initial rotation is
offset by the necessity for a compensatory adjustment in phase of the
passive section rotation although later it will emerge that this is
unimportant for the $N=3$ Deutsch-Jozsa algorithm. This simplification
is akin to working in co-rotating frames (one for each spin) which do
not coincide at the point of the first spin selective rotation and
Eq.~(\ref{eq:ipulses3}) demonstrates that this is legitimate.

\subsection{Linear and quadratic term gate construction}

For linear term gates the resulting idealized pulse sequence is
\begin{equation}
 \left[ U^i_{\mbox{\tiny LIN}} \right] =  \left[ 180^\circ \right]^i_z.
\end{equation}
The function evaluation gate can always be arranged so that the linear
term gates appear after the quadratic term gates. This adds
$180^\circ$ to the passive section ${\bf \hat{z}}$ rotation for spin
$i$ and can be implemented as part of the post-acquisition phase
shift.  This motivates a further classification of admissible
functions based on the similarity of the active sections of the
evaluation step pulse sequence (i.e.\ on the similarity of the
quadratic terms in $f$). Ignoring all linear terms and distinguishing
only between arrangements of quadratic terms, the following classes
emerge: $\{ f_1, f_2, f_3 \}$, $\{ f_4, f_5, f_6 \}$, $\{ f_7, f_8 \}$
and $\{ f_9, f_{10} \}$. Within a given class the pulse sequences
differ only by combinations of $180^\circ$ post-acquisition phase
shifts; their active sections are identical. There are essentially
four different experiments to perform to implement $N=3$ Deutsch-Jozsa
algorithm.

Apart from the initial rotations, the only other contributions to the
active sections of the pulse sequences come from the quadratic term
gates.  Substitution of a controlled-NOT gate pulse sequence,
Eq.~(\ref{eq:cn1}), into the quadratic term gate decomposition,
Eq.~(\ref{eq:quadgateseq}), followed by compilation yields the
idealized pulse sequence
\begin{equation}
 \hat{U}^{ij}_{\mbox{\tiny QUAD}} = \hat{R}^i_{\bf -\hat{z}}(90^\circ) \;
                             \hat{R}^j_{\bf -\hat{z}}(90^\circ) \;
                             \hat{U}^{ij}_{\mbox {\tiny SCAL}}(1/2J_{ij})
\end{equation}
or
\begin{equation}
 \left[ U^{ij}_{\mbox{\tiny QUAD}} \right] = \left[ 1/2J_{ij} \right]^{ij}
              - \left[ 90^\circ \right]^j_{-z}
              - \left[ 90^\circ \right]^i_{-z}.
 \label{eq:quseq}
\end{equation}
The success of a literal translation of this sequence into a
refocused delay followed by relevant phase shifts depends on the
ratio of $\left| J_{ij} \right|$ to the fastest relaxation rate for
the spins which represent the qubits. If this ratio is large enough
then such a {\em directly coupled realization} is feasible.

\begin{figure}[ht]
  \epsfxsize=3in
  \centerline{\epsffile{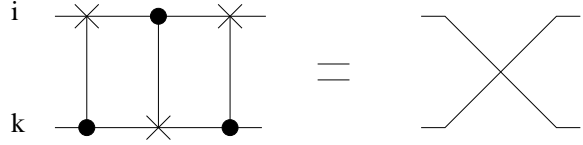}}
  \vspace{2ex}   
  \caption{Decomposition of the SWAP operation,
 $\hat{U}^{ik}_{\textrm{\tiny SWAP}}$, into controlled-NOT gates.}
 \label{pic:swap_cnot}
\end{figure}

\begin{figure}[ht]
  \centerline{\epsffile{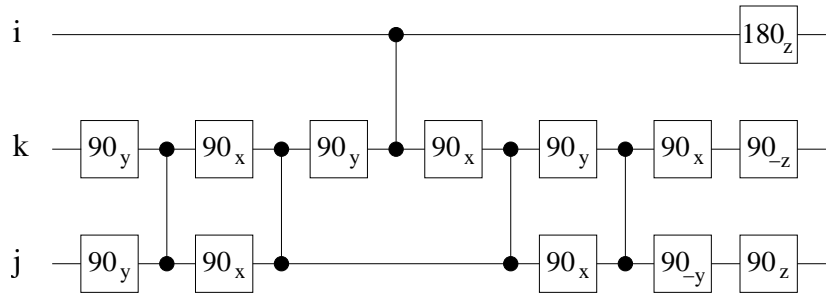}}
  \vspace{2ex}
   \caption{Pulse sequence for indirectly coupled realization of
 $\hat{U}^{ij}_{\textrm{\tiny SCAL}}\left(1/2J_{ij} \right)$
 based on swapping to a controlled-NOT gate.}  \label{pic:quad_cnot}
\end{figure}

On the other hand, if $\left| J_{ij} \right|$ is comparable to or less
than any of the relaxation rates the directly coupled realization will
result in unacceptable errors. However, provided that a network of
substantially stronger couplings connects $i$ and $j$ it may be
possible to implement an {\em indirectly coupled realization} by
swapping information along the coupling pathway. Suppose, for example,
that each of $i$ and $j$ are coupled to $k$ appropriately
strongly. Then the SWAP operation $\hat{U}^{ik}_{\mbox{\tiny SW}}$,
defined on the computational basis as
\begin{equation}
 \hat{U}^{ik}_{\mbox{\tiny SWAP}} \left|x_N \dots x_i  
                                          \dots x_k 
                                          \dots x_0 \right> :=
                                \left|x_N \dots x_k 
                                          \dots x_i 
                                          \dots x_0 \right>
\end{equation}
and extended linearly to all states, allows~\cite{madi} information to
be exchanged via $k$.  A decomposition into controlled-NOT gates is
given in Fig.~\ref{pic:swap_cnot}. 
Two possibles strategies for the
construction of $\hat{U}^{ij}_{\mbox{\tiny SCAL}}(1/2J_{ij})$ are
apparent:

\begin{itemize}
 \item[1.] Use evolution under the coupling between $k$ and $j$ and
the swapping technique to give
\begin{equation}
    \hat{U}^{ij}_{\mbox{\tiny SCAL}}(1/2J_{ij})
                                = \hat{U}^{ik}_{\mbox{\tiny SWAP}} \;
                                  \hat{U}^{jk}_{\mbox{\tiny SCAL}}(1/2J_{jk}) \;
                                  \hat{U}^{ik}_{\mbox{\tiny SWAP}}. 
\end{equation}
One possible compiled pulse sequence is provided in
Fig.~\ref{pic:quad_ind}. Others are available by mixing alternative
constructions for the constituent controlled-NOT gates.
\begin{figure}[ht]
  \centerline{\epsffile{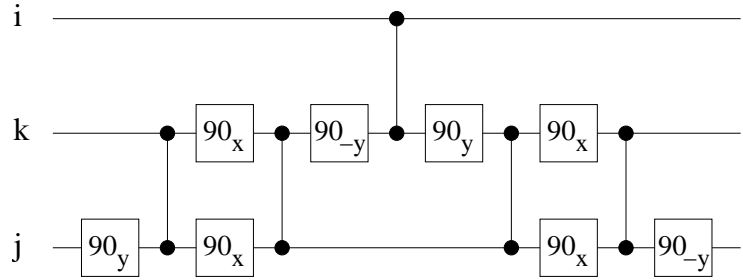}}
  \vspace{2ex}
  \caption{Pulse sequence for indirectly coupled realization of
 $\hat{U}^{ij}_{\textrm{\tiny SCAL}}\left(1/2J_{ij} \right)$
 based on swapping to a scalar coupling gate. Lines connecting qubits
 $i$ and $k$ represent $\hat{U}^{ik}_{\textrm{\tiny
 SCAL}}\left(1/2J_{ik} \right)$. Squares represent single
 qubit rotations.}
\label{pic:quad_ind}
\end{figure}

\item[2.] Use an indirectly coupled controlled-NOT gate between $j$
and $k$ to construct a controlled-NOT gate between $i$ and $j$, which
in turn gives $\hat{U}^{ij}_{\mbox{\tiny SCAL}}(1/2J_{ij})$. Thus
\begin{equation}
\hat{U}^{ij}_{\mbox{\tiny CNOT}} =    \hat{U}^{jk}_{\mbox{\tiny SWAP}} \; 
           \hat{U}^{ik}_{\mbox{\tiny CNOT}} \; 
           \hat{U}^{jk}_{\mbox{\tiny SWAP}}  
\end{equation}
and 
\begin{equation}
 \hat{U}^{ij}_{\mbox{\tiny SCAL}}(1/2J_{ij}) =
                 \hat{R}^j_{\mathbf{-\hat{y}}}(90^\circ) \; 
                 \hat{U}^{jk}_{\mbox{\tiny SWAP}} \;
                 \hat{U}^{ik}_{\mbox{\tiny CNOT}} \; 
                 \hat{U}^{jk}_{\mbox{\tiny SWAP}} \; 
                 \hat{R}^j_{\bf \hat{y}}(90^\circ)
\end{equation}
and simplification gives the sequence of Fig.~\ref{pic:quad_cnot}.
\end{itemize}
The relevant quadratic term gate can then be constructed by appending
the appropriate rotations about $\mathbf{\hat{z}}$.

\subsection{Management of Hamiltonian evolution}

The pulse sequences of Eqs.~(\ref{eq:cn2}) and (\ref{eq:quseq}) or
Figs.~\ref{pic:quad_ind} and \ref{pic:quad_cnot} assume that the
internal Hamiltonian for the molecule, Eq.~(\ref{eq:ham}), is only
active is during the delays which appear in the sequence. Furthermore
during such delays it is assumed that only one scalar coupling term is
active. These assumptions fail in two ways. First, externally applied
spin selective RF pulses have a finite duration, during which the
Hamiltonian is active. Second, during any delays, which are
implemented in practice by allowing the spins to evolve freely for an
appropriate duration, all the terms of the Hamiltonian are active.

During the externally applied pulses this is addressed by neglecting
evolution under the coupling terms, which are typically small in
comparison to the external field strengths, and noting that evolution
under the Zeeman terms can be viewed as rotations about the
$\mathbf{\hat{z}}$ axis.  In this approximation the effect of
Hamiltonian evolution has been to include unwanted qubit-dependent
$\mathbf{\hat{z}}$ rotations.  Fortunately,
Eq.~(\ref{eq:zrot_rearrange}) provides a rule for interchanging any
such $\mathbf{\hat{z}}$ rotation with succeeding rotations about
transverse axes for the same qubit and gives the necessary phase
correction for these. Additionally they commute with any scalar
coupling terms. Thus the Zeeman evolution terms can be interchanged so
that they effectively appear at the end of the algorithm pulse
sequence. This procedure amounts to working in the co-rotating frames,
which essentially remove the Zeeman terms from $\hat{H}$.

The same technique effectively removes the Zeeman terms during delays
for scalar coupling evolution. The remaining difficulty is to ``turn
off'' evolution due to other scalar couplings during the delay. This
is accomplished via standard refocusing schemes~\cite{linden2,leung2,jones3}, which are based on periods of free
evolution under all scalar coupling terms interspersed with qubit
selective $180^\circ$ rotations. For example, the sequence
\begin{equation}
\hat{R}^j_{\mathbf{\hat{m}}}(180^\circ) \; e^{-i\hat{H}T/2} \;
\hat{R}^j_{\mathbf{\hat{m}}}(180^\circ) \; e^{-i\hat{H}T/2} 
\label{eq:refoc_seq}
\end{equation}
where ${\bf \hat{m}} = m_x {\bf \hat{x}} + m_y {\bf \hat{y}}$, results
in evolution of duration $T$ under all scalar couplings \emph{except
those involving spin $j$} together with the identity operation on spin
$i$ (ignoring the Zeeman evolutions as discussed previously).

\subsection{Initialization and readout}

A typical room temperature solution state NMR sample consists of an
ensemble of non-interacting identical molecules. Although the nuclear
spin state of any individual constituent molecule is generally
inaccessible, it is possible to control and acquire information about
the ensemble's average state, typically described in terms of the
system's density operator. Fortunately, it has been possible to modify
conventional quantum computation schemes accordingly. In particular,
initialization schemes, relying on the concept of pseudopure states
have been developed~\cite{chuang2,cory1,cory2} and tomography schemes
allow for the reconstruction of final state density
operators~\cite{chuang2}. The algorithm stage is implemented by
subjecting the ensemble to a pulse sequence derived from the
concatenation of the pulse sequences for the algorithm's constituent
fundamental gates.

Bulk state preparation and tomography schemes typically require
efforts, in the form of additional pulses or repeated experiments,
beyond those demanded by the algorithms. However, it emerges that, for
$N \leq 3$, it is possible to apply the evolution stage (see
Fig.~\ref{pic:djscheme}) directly to a thermal equilibrium initial
state and successfully solve the Deutsch problem with an expectation
value measurement~\cite{linden,zhou}. Analysis of the system state
after the function evaluation step, which is equivalent to running the
algorithm followed by the readout $\hat{R}_{\bf \hat{n}}(90^\circ)$,
demonstrates this for $N=3$ (the reasoning is similar for $N\leq 2$).
Using the product operator formalism, the deviation part of the
thermal equilibrium equilibrium density operator for a weakly coupled
homonuclear NMR system is proportional to~\cite{slichter}
\begin{equation}
  \hat{\rho}_{\textrm{\scriptsize th}} = \hat{I}^2_z + \hat{I}^1_z +
\hat{I}^0_z.
\end{equation}
Setting the initial rotation's axis to ${\bf \hat{n}} = -{\bf
\hat{y}}$ results in
\begin{eqnarray}
 \hat{\rho}_{\textrm{\scriptsize th}} 
             & \stackrel{\hat{R}_{-{\bf \hat{y}}}(90^\circ)}
               {\longrightarrow} &
              -\hat{I}^2_x - \hat{I}^1_x - \hat{I}^0_x \\ 
             & \stackrel{\hat{U}_f}{\longrightarrow} & \hat{\rho}_{f}
\end{eqnarray}
which are listed in Table~\ref{tab:foutput}. The spectrometer measures
the components of the magnetization along the ${\bf \hat{x}}$ and
${\bf \hat{y}}$ axes,
\begin{equation}
\left< M_+ (\varphi)\right>(t) := \sum_{j=0}^2 \mbox{Tr}\, 
                        \left( e^{i\varphi}
                               \left( \hat{I}_x^j + i \hat{I}_y^j \right)\,
                        e^{-i\hat{H}t/\hbar}\,
			\hat{\rho}\,
                        e^{i\hat{H}t/\hbar} \right),
\end{equation}
where $\hat{\rho}$ is the density operator immediately prior to
acquisition. Finally the spectrum resulting from a time domain Fourier
transform of $\left< M_+ (\varphi)\right>(t)$ is calculated and
displayed. The acquisition phase, $\varphi$, describes any overall
phase which is independent of the pulse sequence applied to the
sample. Typically it depends on factors such as the spectrometer's
construction, probe construction or receiver electronics and may be
difficult to determine.  The appearance of any spectrum depends on
$\varphi$, although this can be adjusted arbitrarily by applying phase
shifts to the acquired data. In conventional NMR applications the lack
of knowledge of $\varphi$ does not usually impose any impediments. For
example, post-acquisition phase shifts can be applied to the data so a
line of interest appears in absorption mode. However, for quantum
computation purposes $\varphi$ could be important because the only
possible method of extracting information from $\hat{\rho}$, and hence
results from the computation, is via the acquired spectrum.
Fortunately, the fact that $\varphi$ is constant for different pulse
sequences means that it is completely irrelevant when comparing two
spectra provided that the physical setup (except for the pulse
sequences) used to acquire these is unaltered. In practice such
comparisons are facilitated by adjusting post-acquisition phase
shifts. It follows that the algorithm will have to be answered by
comparing spectra. Signal acquisition {\em immediately after
application of $\hat{U}_f$} and with {\em no additional readout
pulses} provides a spectrum, the {\em $f$-spectrum}, whose form
depends on $f$. This will be compared to a {\em fiducial spectrum}
which is obtained in the same way but with $\hat{U}_f$ replaced by
$\hat{I}$; here the system's pre-acquisition state is
$\rho_{\mbox{\scriptsize fid}} = -\hat{I}^2_x - \hat{I}^1_x -
\hat{I}^0_x$, identical to the $f$-spectrum for a constant function.
The pertinent features of the resulting spectra can be deduced from
the density operator prior to acquisition and depend on the presence
of scalar coupling terms. The analysis is simplest in the case where
all scalar coupling strengths are significantly larger than the line
widths. 
\begin{figure}[ht]
  \epsfxsize=2.2in
  \centerline{\epsffile{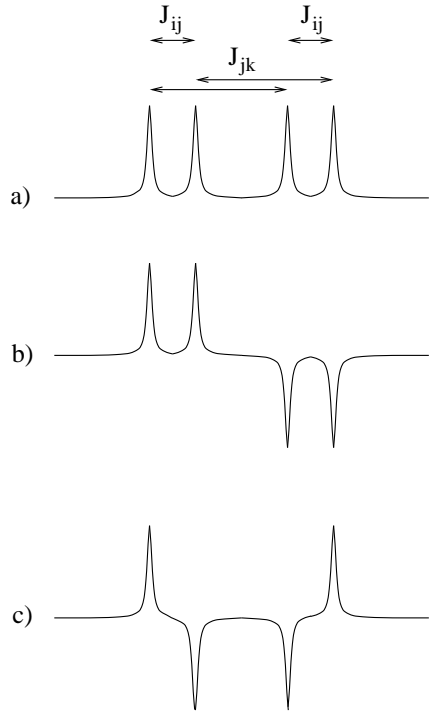}}
  \vspace{2ex}
   \caption{Multiplet forms: a) in-phase quartet arising from $\hat{I}^j_x$, b) antiphase quartet arising from $\hat{I}^i_z \hat{I}^j_x$, and c) doubly antiphase quartet arising from $\hat{I}^i_z \hat{I}^j_x \hat{I}^k_z$.}  \label{pic:multiplets}
\end{figure}
Here each multiplet consists of four distinct lines. The form
of the spin $j$ multiplet (i.e. the lines' relative intensities and
phases) depends on the density operator term containing a factor of
$\hat{I}^j_x$ (see Fig.~\ref{pic:multiplets}). A single factor of
$\hat{I}^j_x$ yields a multiplet with lines of equal phases and
intensities.  $\hat{I}^i_z \hat{I}^j_x$ yields an multiplet in which
the pairs of lines separated by frequency $J_{ij}$ have opposite
phases but equal intensities (an antiphase multiplet).  $\hat{I}^i_z
\hat{I}^j_x \hat{I}^k_z$ yields a multiplet in which pairs of lines
separated by a $J_{ij}$ have opposite phases while simultaneously
those separated by $J_{kj}$ have opposite phases and equal intensities
(a doubly antiphase multiplet). Multiplication of a product operator
term by $-1$ corresponds to a phase shift of $180^\circ$ for the
corresponding multiplet. Now consider the case where one or more of
the scalar couplings is small in comparison with the
linewidths. Suppose, for example, that $J_{ij}$ is small relative to
the linewidths for spin $j$. It follows that the spin $j$ multiplet's
constituent lines which would otherwise be separated by $J_{ij}$ will
overlap. The resulting lineshapes can be predicted by adding two
appropriately overlapping Lorentzian lines. In the case of lines of
opposite phase separated by $J_{ij}$ partial cancellation will
result. In the extreme case, $J_{ij} = 0$, the cancellation is
complete and these lines disappear. Thus the multiplet arising from
any product operator term containing a factor of $\hat{I}^i_z
\hat{I}^j_x$ disappears if $J_{ij} = 0$.

The quantum logic for this algorithm demands that for each qubit is
coupled to at least one other qubit with a scalar coupling much larger
than any relaxation rates (and hence linewidths). For a three qubit
molecule this implies that only one scalar coupling term can be
negligible. In this case Table~\ref{tab:foutput} indicates that at for
$f_7$ to $f_{10}$ at least one antiphase or doubly antiphase multiplet
survives. For $f_4$ to $f_6$ the multiplet for spin $0$ survives but
its lines are all out of phase with those of the fiducial
spectrum. For $f_1$ to $f_3$ the same applies to spin $2$. Similar
results apply for other representatives of the function categories but
with the qubits permuted appropriately. On the other hand the spectrum
for the constant functions is identical to the fiducial spectrum.
Thus we arrive at an algorithm answer criterion for applicable
functions:

(i) {\em $f$ is constant if and only if the $f$-spectrum is identical
 to the fiducial spectrum and}

(ii) {\em $f$ is balanced if and only if there is a $\pi$ phase 
difference between at least one line of the $f$-spectrum and its 
counterpart in the fiducial spectrum.} 

The fiducial spectrum can be phased (effectively adjusting the
acquisition phase) so that its constituent lines all appear
upright. Thus, the answer to the $N=3$ Deutsch problem may be
determined by inspecting the $f$-spectrum for inverted lines. Each
balanced function produces at least one inversion. For constant
functions all lines are upright.  This provides a solution state NMR
scheme for conclusively answering the $N=3$ Deutsch problem with just
one application of the evolution stage to the thermal equilibrium
input state.

\section{Pulse Sequence Construction: Experiment}
\label{sec:exp}

\subsection{Molecular parameters and coupling architecture}

\begin{figure}[ht]
  \epsfxsize=2in
  \centerline{\epsffile{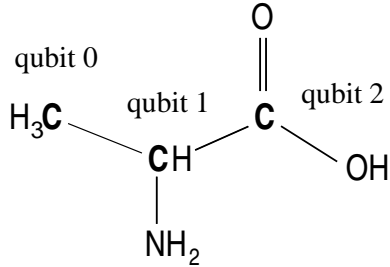}}
  \vspace{2ex}
  \caption{Qubit representation in the alanine molecule.}
  \label{pic:alanine}
\end{figure}

A saturated solution of $^{13}\mbox{C}$ labeled alanine in
$\mbox{D}_2\mbox{O}$ provided the qubits (see
Fig.~\ref{pic:alanine}). We denote the carboxyl carbon, qubit $2$, the
$\alpha$ carbon, 1 and the methyl carbon, $0$. Protons were decoupled
using a standard heteronuclear decoupling technique. Scalar couplings
are $J_{21} = 56\,$Hz, $J_{10} = 36\,$Hz and $J_{20} = 1.57\,$Hz. For
$J_{20}$, the linewidth is comparable to the relevant interpeak
splitting and thus the coupling strength differs from the interpeak
distance.  However, the SWAP method of gate construction allows for a
creation of an antiphase doublet between these peaks without any prior
knowledge of $J_{20}$, giving an accurate method for calculating its
value (see appendix A).  The $T_1$ relaxation times were determined
via an inversion relaxation sequence and $T_2$'s by a single spin
echo.  Results are $T_1(2)=20.3\,$s, $T_1(1)=2.82\,$s,
$T_1(0)=1.45\,$s, $T_2(2)=1.25\,$s, $T_2(1)=0.417\,$s, and
$T_2(0)=0.702\,$s where the argument labels the qubit.

Quadratic term gates between the pair of qubits 1 and 2 or the pair 0
and 1 can be implemented using the directly coupled realization,
Eq.~(\ref{eq:quseq}). Here the refocusing scheme of
Eq.~(\ref{eq:refoc_seq}) suggests
\begin{equation}
 \left[U^{21}_{\mbox{\tiny QUAD}} \right]  = 
                \left[ 1/4J_{21} \right]^{\mbox{\scriptsize tot}}
              - \left[ 180^\circ \right]^0_{n}
              - \left[ 1/4J_{21} \right]^{\mbox{\scriptsize tot}}
              - \left[ 180^\circ \right]^0_{n}
              - \left[ 90^\circ \right]^1_{-z}
              - \left[ 90^\circ \right]^2_{-z}
 \label{eq:u21seq}
\end{equation}
and
\begin{equation}
 \left[U^{10}_{\mbox{\tiny QUAD}} \right]  =  
                \left[ 1/4J_{10} \right]^{\mbox{\scriptsize tot}}
              - \left[ 180^\circ \right]^2_{n}
              - \left[ 1/4J_{10} \right]^{\mbox{\scriptsize tot}}
              - \left[ 180^\circ \right]^2_{n} 
              - \left[ 90^\circ \right]^0_{-z}
              - \left[ 90^\circ \right]^1_{-z}
\end{equation}
where $n$ represents any axis in the $xy$ plane. Note that in each of
these sequences the refocusing scheme eliminates the coupling between
qubits 0 and 2.  Unfortunately, for $\hat{U}^{20}_{\mbox{\tiny
QUAD}}$, which is required for $f_9$ and $f_{10}$, the directly
coupled realization is inadequate since $1/2J_{20} = 0.32\,$s which is
comparable to the smallest $T_2$. However, one of the indirectly
coupled realizations illustrated in Fig.~\ref{pic:quad_ind} or
Fig.~\ref{pic:quad_cnot}, where $i=2, k=1$ and $j=0$, provides a
construction which requires additional pulses but is sufficiently
faster. The approach based on swapping to a controlled-NOT gate (see
Fig.~\ref{pic:quad_cnot}) is preferable despite the fact that it
appears to require more pulses than that based on swapping to a scalar
coupling gate. The reason is that the initial rotation of the
algorithm (see Fig.~\ref{pic:djscheme}) may be chosen with ${\bf
\hat{n}} = -{\bf \hat{y}}$. Then the first two rotations on spins 1
and 0 cancel, leaving the scalar coupling term between 0 and 1 as the
initial operation on these qubits (see
Fig.~\ref{pic:quad_cnot}). However, these are initially in thermal
equilibrium and this operation has no effect and may be omitted. The
resulting pulse sequence has duration $0.056\,$ms, which is significantly
faster than the direct version using $\left[ 1/2J_{20} \right]^{20}$,
and comfortably within the shortest relaxation time.  The pulse
sequence compiles to
\begin{eqnarray} \left[U^{20}_{\mbox{\tiny QUAD}} \right] & =&
 \left[ 90^\circ \right]^0_{x} - \left[ 90^\circ \right]^1_{x} -
 \left[ 1/2J_{10} \right]^{10} - \left[ 90^\circ \right]^1_{y} -
 \left[ 1/2J_{21} \right]^{21} - 
 \nonumber \\ && 
 \left[ 90^\circ \right]^1_{x} -
 \left[ 1/2J_{10} \right]^{10} - \left[ 90^\circ \right]^1_{y} -
 \left[ 90^\circ \right]^0_{x} - \left[ 1/2J_{10} \right]^{10} - 
 \nonumber \\ && 
 \left[ 90^\circ \right]^1_{x} - 
 \left[ 90^\circ \right]^0_{-y} - 
 \left[ 90^\circ \right]^2_{z} - \left[ 90^\circ \right]^1_{-z}
 \label{eq:f9seq}  
\end{eqnarray}
where
\begin{equation}
 \left[1/2J_{21} \right]^{21}  = 
                \left[ 1/4J_{21} \right]^{\mbox{\scriptsize tot}}
              - \left[ 180^\circ \right]^0_{n}
              - \left[ 1/4J_{21} \right]^{\mbox{\scriptsize tot}}
              - \left[ 180^\circ \right]^0_{n}
 \label{eq:realrefoc1} 
\end{equation}
and
\begin{equation}
 \left[1/2J_{10} \right]^{10}  =  
                \left[ 1/4J_{10} \right]^{\mbox{\scriptsize tot}}
              - \left[ 180^\circ \right]^2_{n}
              - \left[ 1/4J_{10} \right]^{\mbox{\scriptsize tot}}
              - \left[ 180^\circ \right]^2_{n}. 
 \label{eq:realrefoc2} 
\end{equation}
Again the refocusing schemes eliminate the coupling between qubits $0$
and $2$. Thus, only the linear coupling chain $0-1-2$ is used to carry out
the algorithm's evolution stage; for all admissible functions the
value of $J_{20}$ is irrelevant during this step.

The pulse sequences for $f_9$ are obtained by concatenating those of
Eqs.~(\ref{eq:u21seq})~-~(\ref{eq:f9seq}).

\subsection{Pulse parameterization}
 
The experiments were performed at room temperature using a Bruker
500-DRX spectrometer and an inverse detection probe. The bulk of the
experimental effort was devoted to constructing
$\left[U^{20}_{\mbox{\tiny QUAD}} \right]$ according to
Eq.~(\ref{eq:f9seq}) and consequently to implementing the algorithm
for $f_9$ and $f_{10}$.

All realizations required qubit selective $90^\circ$ rotations and
most also used selective $180^\circ$ rotations for refocusing. Shaped
Gaussian pulses with $10\, \%$ truncation and durations of $0.7$ms for
qubits 0 and 1 and duration $0.5$ms for qubit 2 gave sufficient
selectivity. No hard pulses were used. The power parameters for
selective rotations were found by applying the selective pulse to the
thermal equilibrium state followed by signal acquisition; this
procedure was repeated, starting at a low power and increasing this
with each application. The resulting acquired signals with maximal and
minimal intensities correspond to $90^\circ$ and $180^\circ$
rotations. The effectiveness of the resulting $90^\circ$ rotation can
be examined partially by applying the following sequence to the
thermal equilibrium state:

\begin{equation}
 \left[ 90^\circ \right]^i_{\bf \hat{y}} - \mbox{delay } T -
 \left[90^\circ \right]^i_{\cos{\theta}{\bf \hat{x}}
                          + \sin{\theta}{\bf\hat{y}}} - \mbox{acquire}
 \label{eq:rotTrot}
\end{equation}

where $T$ is small enough to neglect any scalar coupling effects. For
a rotation through exactly $90^\circ$ the acquired signal's intensity
is proportional to $\cos{(2\pi\nu T)}$ where $\nu$ is the offset
relative to the receiver frequency. Furthermore, if the axis of the
second rotation compared to that of the receiver remains constant as
$T$ varies, then the phase of the acquired spectrum should remain
constant. We found that this was resulting signal dependence on $T$,
particularly in the region where the signal intensity approaches a
minimum, is fairly sensitive to variations in the pulse power. We used
this to perform fine adjustments to the power levels for $90^\circ$
pulses.

Construction of the real pulse sequence corresponding to
Eq.~(\ref{eq:f9seq}) is accomplished by supplying the correct
parameters for the active section of the pulse sequence (such as pulse
phases and duration of delays) followed by computation of the correct
passive section parameters. The division of the pulse sequence into
active and passive parts is reflected in the different techniques
required for realizing the correct operations.

\subsection{Passive section parameterization}

The passive section of the algorithm's pulse sequence (see
Eq.~(\ref{eq:passive})), is implemented by adjusting qubit-dependent
phase shifts after signal acquisition. For the
fiducial case and any $f$ the passive section parameter contains
absolute phase information (unlike the relative rotation axes of the
active section) given by the algorithm and assorted corrections which
arise from Zeeman evolution during the active section. However, the
fact that the algorithm answer criterion is based on comparing
$f$-spectra line phases to those in the fiducial spectrum means that
for any $f$ only $\phi_f^i - \phi_{\mbox{\scriptsize fid}}^i$ are
important where $\phi_f^i$ are the passive section parameters for $f$
and $\phi_{\mbox{\scriptsize fid}}^i$ those for the fiducial
case. Such relative values are determined by the algorithm and are
independent of the acquisition phase and choice of the arbitrary
initial rotation phases (provided that these remain constant for all
admissible functions and the fiducial case). It follows that an
acceptable procedure is to begin by acquiring the fiducial spectrum,
phase this to give upright absorption-mode multiplets and save the
resulting post-acquisition phase shifts. Then, for any $f$, the active
section of the algorithm pulse sequence is applied followed by
post-acquisition phase shifts, $\phi_f^j - \phi_{\mbox{\scriptsize
fid}}^j$, added to the values stored after processing of the fiducial
spectrum. The resulting output spectra will be phased in accordance
with those of Fig.~\ref{pic:multiplets}.

\subsection{Active section  parameterization}

The active section of the algorithm (see Eq.~(\ref{eq:active})) is
implemented by applying a sequence of spin selective Gaussian shaped
pulses with the appropriate rotation axes and refocused delays of the
appropriate durations. These parameters are derived from the
corresponding algorithm sequence; for example, those for $f_9$ are
derived from Eq.~(\ref{eq:f9seq}). Applying a pulse program with
rotation axes and delays taken literally from Eq.~(\ref{eq:f9seq}) is
unsatisfactory because this would ignore effects such as the ongoing
Zeeman evolution discussed earlier. In fact, the bulk of the effort is
devoted to tuning these parameters correctly.

Axes of rotation are determined by pulse phases, which can be
specified with a resolution of $360^\circ/65536 = 0.0054932^\circ$ on
the Bruker equipment. In general it is not straightforward to
associate spectrometer phases with absolute rotation axes such as
those required by Eq.~(\ref{eq:f9seq}). However,
Eq.~(\ref{eq:ipulses3}) shows that it is only the relative axes and
hence relative pulse phases which are important. The primary
difficulty in adjusting these relative phases correctly arises from
evolution under the Zeeman terms during pulse applications and
delays. There are three cases:

\begin{itemize}
 \item[i)] Zeeman evolution during delays,
 \item[ii)] Zeeman evolution experienced by spectator qubits during
application of any selective pulse and
 \item[iii)] Zeeman evolution experienced by the rotated qubit during
a selective pulse.
\end{itemize}

During a delay of duration $t$ the Zeeman terms produce
qubit-dependent rotations about ${\bf \hat{z}}$ through angles $2 \pi
\nu^j t$ where $\nu^j$ is the offset frequency of qubit $j$. This is
corrected by re-expressing the idealized delay term as
\begin{equation}
 \left[ t  \right]^{\mbox{\scriptsize tot}} 
  = \left[ t \right]^{\mbox{\scriptsize tot}}
   - \left[ 2 \pi \nu^2 t \right]^2_z
   - \left[ 2 \pi \nu^1 t \right]^1_z
   - \left[ 2 \pi \nu^0 t \right]^0_z
   - \left[ -2 \pi \nu^1 t\right]^2_z
   - \left[ -2 \pi \nu^0 t  \right]^1_z
   - \left[ -2 \pi \nu^0 t \right]^0_z
\end{equation}
and noting that if the Zeeman terms are correctly included then the
spectrometer actually implements $\left[ t \right]^{\mbox{\scriptsize
tot}}-\left[ 2 \pi \nu^2 t \right]^2_z - \left[ 2 \pi \nu^1 t
\right]^1_z - \left[ 2 \pi \nu^0 t \right]^0_z.$ The corrections
$\left[ -2 \pi \nu^2 t \right]^2_z - \left[ -2 \pi \nu^1 t \right]^1_z
- \left[ -2 \pi \nu^0 t \right]^0_z$ are then rearranged according to
the compilation scheme with the result that for qubit $j$ the axes of
rotation of succeeding rotations must be shifted through $2 \pi \nu^j
t$ and the final phase shift must be altered according to $\phi^j
\rightarrow \phi^j - 2 \pi \nu^j t.$

During a selective pulse of duration $t$ to any qubit the same rules
apply to spectator qubits. The remaining complication is the effect of
Zeeman terms on the qubit to which the selective pulse is
applied. These can be calculated analytically for rectangularly shaped
pulses by passing to a rotating frame; the results are similar to
those for delays~\cite{slichter}. However, to the best of our
knowledge, no such comparable result exists for Gaussian shaped
pulses. Our approach is to adjust the phases experimentally by
considering the form of the spectrum acquired after each spin
selective rotation. This sometimes requires readout pulses,
judiciously selected to reveal key features of the density operator
without going to the extreme of complete state tomography. This
technique depends on a complete knowledge of the density operator
prior to acquisition. Although this is not satisfactory for general
purpose quantum computation, it is sufficient to demonstrate the
essential dynamics. Furthermore it should be possible to simulate the
exact effects of simultaneously Zeeman evolution and rotation under
Gaussian shaped pulses.

In principle, evolution under the scalar coupling terms also provides
erroneous effects during the pulse sequence. Fortunately, the $\left[
1/4J_{ij} \right]^{ij}$ delays are exempt from such errors since these
are designed to utilize these couplings. It remains to assess the
effects during $5\,\mu$s delays between selective pulses (required by
the spectrometer for switching purposes) and during the selective
pulses. The extent of such evolution can be estimated crudely by
comparing $t$ to $1/2J_{ij}$, which represents the ``maximum'' evolution
under the relevant scalar coupling. Thus a crude measure of the
fractional error is $t /\frac{1}{2J_{ij}} = t 2 J_{ij}$ where $t$ is the
duration of the delay or pulse. For a $5\,\mu$s delay this yield at
most $0.00056$ and for the $0.7\,$ms pulses and the strongest
coupling, $0.078.$ Thus to a fair approximation the scalar coupling
terms can be neglected during selective pulses.

The remaining experimental issue is to construct the refocused delays
of Eqs.~(\ref{eq:realrefoc1}) and~(\ref{eq:realrefoc2}). The durations
were determined experimentally by comparing the signal acquired after
the delay to that calculated theoretically. For $\left[ 1/4J_{10}
\right]^{\mbox{\scriptsize tot}}$ the optimal delay was found to be
$6.5\,$ms and for $\left[ 1/4J_{21} \right]^{\mbox{\scriptsize tot}}$
it was $3.95\,$ms. The respective theoretical values were $6.9\,$ms
and $4.5\,$ms. The discrepancies can only be partially explained by
coupling evolution during the $1.4\,$ms required for the refocusing
pulses. In principle, the phases of the two $180^\circ$ pulses within
any refocused delay relative to those elsewhere in the pulse sequence
are unimportant; only the relative phases of the two matter. These
were again determined experimentally by the general process described
for the selective pulses. An additional improvement is attained by
adding the results of two experiments in which the sequences are
identical except that in the second the axes of the refocusing pulses
were both offset by $180^\circ$. This is done using the spectrometer's
phase cycling routines. In theory the pulse sequences are identical;
in practice the refocusing appears to be more accurate using this
scheme.

\subsection{Spectral output and results}

\begin{figure}[ht]
  \epsfxsize=3.375in
  \centerline{\epsffile{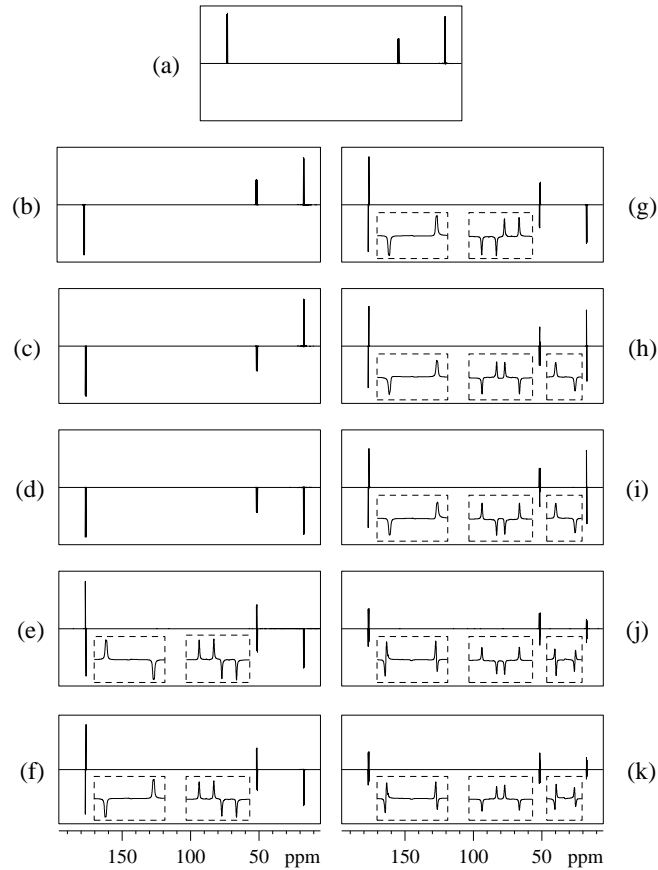}}
  \vspace{2ex}
   \caption{$\mbox{}^{13}$C output spectra for alanine: (a) Fiducial
spectrum, (b) $f_1$ spectrum, (c)$f_42$ spectrum, (d) $f_3$ spectrum,
(e) $f_4$ spectrum, (f) $f_5$ spectrum, (g) $f_6$ spectrum, (h) $f_7$
spectrum, (i) $f_8$ spectrum, (j) $f_9$ spectrum, and (k) $f_{10}$
spectrum. In each spectrum the leftmost multiplet corresponds to qubit
2, the central to qubit 1 and the rightmost to qubit 0. Insets provide
enlargements of all antiphase and doubly antiphase multiplets.}
\label{pic:output}
\end{figure}

For each representative function listed in Table~\ref{tab:foutput}
signal acquisition takes place immediately after implementation of
$\hat{U}_f$. Figure~\ref{pic:output} provides selected experimental
spectra that are phased so that $-\hat{I}_x^j$ product operator terms
correspond to upright multiplets. In general it is necessary to adjust
the post-acquisition phase for each multiplet. The Bruker-500
spectrometer allows only two adjustments, a constant and linear phase
shift, per processed spectrum. The entire spectrum can be displayed
with the correct phases by inserting, immediately before acquisition,
a qubit selective $180^\circ$ rotation about the appropriate axis in
the $xy$ plane for the remaining multiplet. If the axis of rotation is
chosen correctly relative to the density operator terms for the qubit
then it has the same effect as a qubit-dependent phase shift. This
requires knowledge of the final density operator and is not useful for
general purpose quantum computation. In our experiment it is merely
used for convenient display of the $f$-spectra. The arrangements of
in-phase, antiphase and doubly antiphase multiplets agree with those
predicted from $\hat{\rho}_f$, indicating that the correct dynamical
evolution for the algorithm had been implemented. An interesting
feature is that the doubly antiphase multiplets for qubits 0 and 2 are
well resolved in the $f_9$ and $f_{10}$ cases (see
Fig.~\ref{pic:output}(e)). For qubit 2 the small $J_{20}$ as well as
the larger $J_{21}$ splitting are clearly discernable. Similarly for
qubit $0$. The improvement in resolution of the $J_{20}$ splitting in
comparison to the inphase case stems from the fact that the resolution
of an antiphase doublet is always better than that of an inphase
doublet whenever the linewidth is comparable to the splitting (see
appendix A). It follows that this indirectly coupled gate realization
offers a method for determining the values of small couplings.

An estimate of errors for the most complicated case, $f_9$, may be
obtained by applying a selective $90^\circ$ readout pulse about the
$\hat{\bf x}$ axis immediately after $\hat{U}_f$. Ideally the readout
spin multiplet should remain while the others disappear.  The average
amplitudes of the residual signals for the latter lie between $14\%$
and $31\%$ of the average amplitude of the corresponding lines with no
readout. The ability to extract the Deutsch problem solution for $N
\leq 3$ via pure phase information in the output spectrum, in contrast
to amplitude information, has clearly mitigated such errors. The most
likely source of error are imperfections in the selective
rotations. The most complicated case, $f_9$, required eight selective
$90^\circ$ rotations, during each of which possible effects of scalar
coupling evolution during application were ignored. Indeed, for the
indirectly coupled realization of $\hat{U}^{20}$ the total duration of
all eight selective rotations, $5.6\,$ms, is $63\%$ of $1/2J_{21} =
8.9\,$ms. To the best of our knowledge there is no satisfactory
analytical understanding of the effects of coupling dynamics during
selective pulses. At the very least such effects need to be modeled in
terms of typical quantum information processing operations and must be
applicable to shaped selective pulses. 

This work was supported, in part, by the DARPA and the ONR. We would
 also like to thank Gary Sanders for useful discussion.

\section{Appendix A}

Lineshapes are predicted via $x$ and $y$ magnetization components,
which are solutions of the Bloch equations,~\cite{harris}
\begin{eqnarray}
 M_x(\Delta \omega) & = & \omega_1 T^2_2 M_0  \frac{\Delta\omega}
                 {1 + T^2_1T^2_2\omega_1^2 + T^2_2(\Delta\omega)^2 } \\
 M_y(\Delta \omega) & = & \omega_1 T^2_2 M_0 \frac{1}
                 {1 + T^2_1T^2_2\omega_1^2 + T^2_2(\Delta\omega)^2 } 
\end{eqnarray}
where $\Delta \omega$ is the offset from the resonance frequency,
$\omega_1= \gamma B_1$ where $B_1$ is the perturbing field strength,
$M_0$ the magnetization after relaxation and $T_1$ and $T_2$ are the
relaxation times. For weak perturbations $T^2_1T^2_2\omega_1^2 \ll 1$
\begin{eqnarray}
 M_x(\Delta \omega) & = &  \frac{2 A}
                 {1 + 4 (\Delta\omega/\Delta\omega_h)^2} \\
 M_y(\Delta \omega) & = &  \frac{A \Delta\omega/\omega_h}
                 {1 + 4 (\Delta\omega/\Delta\omega_h)^2 } 
\end{eqnarray}
where $A=\omega_1 T^2_2 M_0$ is the maximum amplitude attained by
either magnetization and $\Delta\omega_h = 2/T_2$ is the half width of
$M_y$. By setting 
\begin{equation}
 u:= \Delta\omega/\Delta\omega_h
\end{equation}
and 
\begin{equation}
 m_{x,y} := M_{x,y}/A
\end{equation}
then
\begin{eqnarray}
 m_x(u) & = &  \frac{2u}
                 {1 + 4 u^2} \\
 m_y(u) & = &  \frac{u}
                 {1 + 4 u^2} 
\end{eqnarray}
describe the lineshapes in frequency units of $\omega_h$ and
magnetization units of $A$. Suppose that two lines with resonance
(angular) frequencies separated by $2\pi J$ are superimposed. In units
of linewidth the separation is
\begin{equation}
s:= (2\pi J)/\Delta\omega_h.
\end{equation}
Thus an {\em inphase doublet in absorption mode} is described by
\begin{equation}
  m(u) = \frac{2u}{1 + 4 (u-s/2)^2} + \frac{2u}{1 + 4 (u+s/2)^2}
\end{equation}
It is straightforward to show that this has three distinct extrema
whenever $s>1/\sqrt{3}$ and only one when $s \leq 1/\sqrt{3}$. In the
former case two maxima (peaks) of height $1/2s^2(\sqrt{1+1/s^2} -1)$
are separated by $\sqrt{2s\sqrt{1+s^2} - (1+s^2)}$ and are equidistant
from a minimum (interpeak trough) of height $2/(1+s^2)$. In the latter
case the two formerly distinct peaks collapse into a single peak of
height $2/(1+s^2)$. 

On the other hand an {\em antiphase doublet in absorption mode} is
described by
\begin{equation}
  m(u) = \frac{2u}{1 + 4 (u-s/2)^2} - \frac{2u}{1 + 4 (u+s/2)^2}
\end{equation}
For all $s>0$ this has two extrema; a maximum and minimum of equal
amplitude and which are separated by
\begin{equation}
 2\sqrt{s^2-1+\sqrt{s^4+s^2+1}}/(2\sqrt{3}).
 \label{eq:antisep}
\end{equation}
Thus for $0 < s\leq1/\sqrt{3}$ the two peaks of the antiphase doublets
will, in principle, always be distinct whereas those of the inphase
doublet always collapse into a single peak.

It may seem that this would provide a method for determining $J$ based
on the peak separation. However, unless $\Delta \omega_h$ is known, 
this is impossible as Eq.~(\ref{eq:antisep}) only applies when the
separation is given in units of linewidth and the spectrometer only
supplies such measurements in units of Hz. Nor would it be possible to
extract such information from the peak heights as these require $A$,
another typically unknown variable.

A convenient way around these difficulties is to phase the antiphase
doublet so that the lines are in {\em dispersion mode}. Here the
spectrum is described by
\begin{equation}
  m(u) = \frac{2u -s}{1 + 4 (u-s/2)^2} - \frac{2u +s}{1 + 4 (u+s/2)^2}
\end{equation}
The number and arrangement of extrema depends on $s$ but for
$s<\sqrt{3}$, which is the region of most interest, there are two
maxima of equal intensity equally spaced from a single minimum. The
maxima and minima are
\begin{eqnarray}
 m_{\mbox{\scriptsize max}} & = & \frac{s}{2(1+\sqrt{1+s^2})} \\
 m_{\mbox{\scriptsize min}} & = & -\frac{2s}{1+s^2}.
\end{eqnarray}
The two roots of $m$ are situated between the maxima and are separated by 
\begin{equation}
 \Delta u_0 = \sqrt{1+s^2}.
 \label{eq:roots}
\end{equation}
The two maxima, which are easily identifiable, are separated by 
\begin{equation}
 \Delta u_{\mbox{\scriptsize max}} = \sqrt{s^2+1 + 2 \sqrt{1+s^2}}.
 \label{eq:maxsplit}
\end{equation}
The ratio
\begin{equation}
 r_a := \frac{m_{\mbox{\scriptsize max}}}{m_{\mbox{\scriptsize min}}} = 
        \frac{1+s^2}{4(1 + \sqrt{1+s^2})}
 \label{eq:ampratio}
\end{equation}
is independent of the units in which the peak amplitudes are measured
and can be determined from the spectrometer output. Inversion of
Eq.~(\ref{eq:ampratio}) equation yields $s$. Then,
Eq.~(\ref{eq:maxsplit}) gives $\Delta u_{\mbox{\scriptsize max}}$ (in
units of linewidth). This is related to the value determined from the
output spectrum (in units of Hz) by a factor of $\Delta\omega_h$. This
gives the linewidth, $\Delta\omega_h$, which together with $s$ gives
$J$. It is important to note that all that this method requires is the
ability to create an antiphase doublet. For weak scalar couplings it
is possible to do this without any knowledge of $J$ provided that a
network of strong couplings connects the relevant spins.

\end{document}